\documentclass{emulateapj}
\usepackage{apjfonts}
\usepackage{longtable}

\slugcomment{Accepted for publication in ApJS}
\shortauthors{Lim et al.}
\shorttitle{Low-Resolution Spectroscopy for the GCs with Signs of SN enrichment}

\begin{document}

\title{Low-Resolution Spectroscopy for the Globular Clusters with Signs of Supernova Enrichment: M22, NGC 1851 and NGC 288}

\author{
Dongwook Lim\altaffilmark{1,2,3},
Sang-Il Han\altaffilmark{1,2},
Young-Wook Lee\altaffilmark{2,3,8},
Dong-Goo Roh\altaffilmark{2,4}, 
Young-Jong Sohn\altaffilmark{3},
Sang-Hyun Chun\altaffilmark{5},
Jae-Woo Lee\altaffilmark{6},
and Christian I. Johnson\altaffilmark{7}
}

\altaffiltext{1}{Both authors have contributed equally to this paper}
\altaffiltext{2}{Center for Galaxy Evolution Research, Yonsei University, Seoul 120-749, Korea}
\altaffiltext{3}{Department of Astronomy, Yonsei University, Seoul 120-749, Korea}
\altaffiltext{4}{Korea Astronomy and Space Science Institute (KASI), Daejeon 305-348, Korea}
\altaffiltext{5}{Yonsei University Observatory, Seoul, 120-749, Korea}
\altaffiltext{6}{Department of Astronomy and Space Science, Sejong University, Seoul 143-747, Korea}
\altaffiltext{7}{Harvard-Smithsonian Center for Astrophysics, 60, Garden Street, MS-15, Cambridge, MA 02138, USA}
\altaffiltext{8}{Corresponding author: ywlee2@yonsei.ac.kr}

\begin{abstract}
There is increasing evidence for the presence of multiple red giant branches (RGBs) in the color-magnitude diagrams of massive globular clusters (GCs).
In order to investigate the origin of this split on the RGB, we have performed new narrow-band Ca photometry and low-resolution spectroscopy for M22, NGC 1851, and NGC 288.
We find significant differences (more than 4$\sigma$) in calcium abundance from the spectroscopic HK$'$ index for M22 and NGC 1851.
We also find more than 8$\sigma$ differences in CN band strength between the Ca-strong and Ca-weak subpopulations for these GCs.
For NGC 288, however, a large difference is detected only in the CN strength.
The calcium abundances of RGB stars in this GC are identical to within the errors.
This is consistent with the conclusion from our new Ca photometry, where the RGB splits are confirmed in M22 and NGC 1851, but not in NGC 288.
We also find interesting differences in CN-CH correlations among these GCs.
While CN and CH are anti-correlated in NGC 288, they show positive correlation in M22.
NGC 1851, however, shows no difference in CH between the two groups of stars with different CN strengths.
We suggest that all of these systematic differences would be best explained by how strongly type II supernovae enrichment has contributed to the chemical evolution of these GCs.
\end{abstract}


% Section 1 (Introduction) 
\section{INTRODUCTION}
\label{sec_introduction}
Recent studies of stellar populations in globular clusters (GCs) are facing a new paradigm as more and more GCs are observed to have multiple populations.
Photometric observations discovered, for many GCs, splits from main sequence (MS) to red giant branch (RGB) on the color-magnitude diagram (CMD) \citep[e.g.,][]{Lee99,Pan00,Bed04,And09,Han09,jwlee09a,jwlee09b,Bel13,Mil13,Pio13}.
These observations are generally interpreted to be due to the differences in light/heavy elements and/or helium abundances between the subpopulations in these GCs \citep{Nor04,Lee05,Pio07,Cas08,JL13}.
High and medium resolution spectroscopic observations have also established that abundance variations in light elements (C, N, O, Na, Mg and Al) are common in most GCs  \citep[][and references therein]{Gra12a}.
Furthermore, star-to-star variations in heavy elements abundance are discovered in some massive GCs, such as $\omega$~Centauri, M2, M22, M54, NGC 1851, and NGC 2419 \citep{Nor96,Da09,Car10a,Car10b,Car11b,Coh10,JP10,Mar09,Mar11a,Mar11b,Muc12,Yong14b}.

Abundance variations in light elements are suggested to be due to the chemical pollutions and/or enrichments by intermediate-mass asymptotic giant branch (IMAGB) stars \citep{VD08}, rotating AGB stars \citep{Dec09}, and fast-rotating massive stars (FRMSs; \citealt{Dec07}).
However, heavy element abundance variations would imply that a later generation of stars was enriched by type II supernovae (SNe; \citealt{Tim95}).
This in turn indicates that the GCs were once massive enough to retain SNe ejecta, and may even be the remnant cores of dwarf galaxies.
Therefore, searches for these GCs are important as they can provide additional evidence for the Galaxy building blocks predicted in the {$\Lambda$}CDM hierarchical merging paradigm.

In this respect, the narrow-band Ca photometry, which measures the strength of calcium II H \& K lines \citep{Ant91}, can be a powerful probe as it can detect even a small spread in calcium abundance.
Indeed, recent Ca photometry performed at Cerro Tololo Inter-American Observatory (CTIO) has successfully detected splits on the RGB in several GCs, including M22, NGC 1851, and NGC 288 \citep{jwlee09a, jwlee09b, Roh11}.
The caveat in such a narrow-band study, however, is that the adjacent CN lines at 3883 {\AA} can contaminate the measurement if the filter transmission function intrudes into the CN band.
In fact, \citet{Lee13} and \citet{Hsyu14} reported that this is the case in the Ca filter used by previous investigators at CTIO.
Therefore, new photometry employing a new Ca filter, which is carefully designed to avoid this CN contamination, together with follow-up multi-object spectroscopy, are needed to confirm the origin of the split discovered from previous photometric studies.
The purpose of this paper is to confirm the presence and origin of multiple RGBs in M22, NGC 1851, and NGC 288 from the new Ca photometry and low-resolution spectroscopy. As a comparison, we have also observed NGC 6397, which shows only a single narrow RGB \citep{jwlee09a}.

% Section 2 (Narrow-Band Ca Photometry) 
\section{Narrow-Band Ca Photometry}
\label{sec_photometry}
\begin{figure}
\centering
\includegraphics[width=0.48\textwidth]{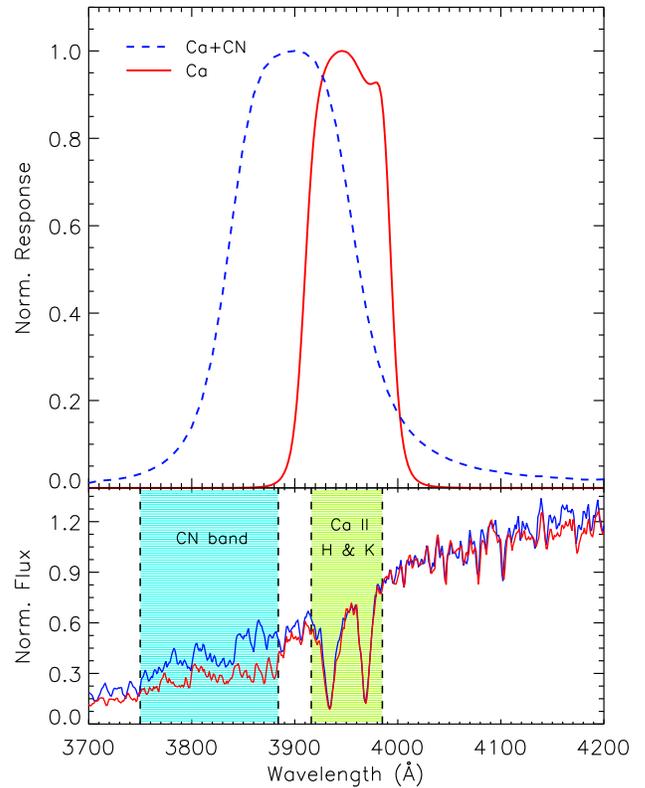}
\figcaption{Comparison of two filter response functions.
Upper panel: Transmission function of old calcium filter (Ca+CN filter) available at CTIO (blue dashed line), which has apparently deteriorated and shifted to blue by 75 {\AA} during the last $\sim$20 years, is compared with that of our new calcium filter (Ca filter) designed to avoid the CN contamination (red solid line).
Lower panel: Two example spectra for CN-normal (blue line) and CN-strong (red line) RGB stars are also compared to illustrate the wavelength regime covered by CN band.
\label{fig_cafilters}}
\end{figure}

The original Ca filter, as defined by \citet{Ant91}, was centered at 3995 {\AA} to include calcium II H \& K lines with a full width half maximum (FWHM) of approximately 90 {\AA}.
Its lower limit was designed to avoid contamination by the CN band at 3883 {\AA}.
In practice, however, the response function of Ca filter might be deteriorated as a result of ageing.
Recently, upon our request, new measurement for the response function of the ``old'' Ca filter available at CTIO was made by C. Johnson, D. H\"{o}lck, \& A. Kunder (2012, private communication).
As briefly reported in \citet{Lee13} and \citet{Hsyu14}, it shows that the response function is shifted to short wavelength by $\sim$75 {\AA} compared to that known to the community.
The origin of this deterioration is not clear, but this underscores the importance of routine inspection of the response function when narrow-band filters are used.
Therefore, we have carefully designed and manufactured a ``new'' Ca filter to avoid the CN contamination.
The blue end of the new filter is shifted to the longer wavelength and has a narrower FWHM (84 {\AA}) with steeper blue and red wings than those of the old Ca filter (see Figure~\ref{fig_cafilters}).
Therefore, while the old Ca filter (hereafter Ca+CN filter) available at CTIO was measuring not only calcium II H \& K lines but also CN band, the new Ca filter (hereafter Ca filter) is practically measuring only calcium lines.

%%%%%%%%%%%%%%%%%%%%%%%%%%%%%%% Figure 1 'Ca filter' %%%%%%%%%%%%%%%%%%%%%%%%%%%%%%%%%%%
%%%%%%%%%%%%%%%%%%%%%%%%%%%%%%%%%%%%%%%%%%%%%%%%%%%%%%%%%%%%%%%%%%%%%%%%%%%%%%%%

Our photometry with the new Ca filter on the du Pont 2.5m telescope at Las Campanas Observatory (LCO) was made during the nights from 2011 to 2013.
The direct CCD equipped on this telescope provides a field of view (FOV) of 8.85{\arcmin} $\times$ 8.85{\arcmin} and a pixel scale of 0.259{\arcsec}.
In order to compare with the data obtained from new Ca filter, we have also reduced the Ca+CN filter data from the CTIO 4m telescope.
This photometry employed the same Ca+CN filter originally used by \citet{jwlee09a,jwlee09b} in their pioneering study.
The observation log for the program GCs is presented in Table~\ref{tab_photlog}.
Following the usual manner \citep[see][]{Han09,Roh11,Lee13}, the raw data were pre-processed by CCDRED package in IRAF\footnote{IRAF is distributed by the National Optical Astronomy Observatory, which is operated by the Association of Universities for Research in Astronomy (AURA) under cooperative agreement with the National Science Foundation.}.
All data were then reduced using DAOPHOT II/ALLSTAR and ALLFRAME \citep{Ste87,Ste90,Ste94}.
Standard stars in M22, NGC 3680, and NGC 5822 \citep{Ant04,jwlee09a,Carraro11} were observed to calibrate instrumental magnitudes to the standard system.
The astrometric solutions were derived from the reference stars in the USNO-B1 and 2MASS All-Sky Point Source catalogs by running tasks in the IRAF FINDER package.

%%%%%%%%%%%%%%%%%%%%%%%%%%%%%%% Table 'log.photometry' %%%%%%%%%%%%%%%%%%%%%%%%%%%%%%%%%%%%
\begin{deluxetable*}{llllll}[hb!]
\tabletypesize{\normalsize}
\tablewidth{0pt}
\tablecaption{Photometric Observation Log}
\tablehead{
\colhead{Observatory}   & \colhead{Object}     & \colhead{Date}     & \multicolumn{3}{c}{Exposures (N$\times$s)}  \\
\colhead{}   & \colhead{}   &   \colhead{}   & \colhead{$y$}        &   \colhead{$b$}   &  \colhead{$Ca$}
}
\startdata
  LCO  &  M22     & 2011 Jun 24 -- 26 & 3$\times$15  & 3$\times$30   &   3$\times$150    \\
       &          &                   & 3$\times$90  & 3$\times$180  &   3$\times$900    \\
       &          & 2013 Jun 11 -- 15 & 3$\times$60  & 3$\times$120  &   3$\times$600    \\
       &NGC~1851  & 2011 Feb 13 -- 14 & 2$\times$30  & 2$\times$60   &   2$\times$300    \\
       &          &                   & 3$\times$180 & 3$\times$360  &   3$\times$1800   \\
       &          & 2012 Feb 13 -- 17 & 5$\times$30  & 5$\times$60   &   5$\times$300    \\
       &          &                   & 5$\times$180 & 5$\times$360  &   5$\times$1800   \\
       &          & 2013 Mar 21 -- 23 & 8$\times$30  & 8$\times$60   &   8$\times$300    \\
       &NGC~288   & 2011 Jun 25 -- 27 & 2$\times$15  & 2$\times$30   &   2$\times$150    \\
       &          &                   & 2$\times$90  & 2$\times$180  &   2$\times$900    \\
       &          & 2013 Jun 15       & 1$\times$15  & 1$\times$30   &   1$\times$150    \\
       &          &                   & 1$\times$90  & 1$\times$180  &   1$\times$900    \\
\hline
  CTIO &  M22     & 2009 Jul 29 -- 30 & 2$\times$20  & 2$\times$40   &   2$\times$240    \\
       &          &                   & 1$\times$200 & 1$\times$400  &   1$\times$1800   \\
       &NGC~1851  & 2009 Jul 29       & 1$\times$25  & 1$\times$50   &   2$\times$600    \\
       &          &                   & 1$\times$200 & 1$\times$400  &   1$\times$900    \\
       &NGC~288   & 2009 Jul 28       & 1$\times$12  & 1$\times$24   &   1$\times$150    \\
       &          &                   & 1$\times$120 & 1$\times$240  &   1$\times$1500
\enddata
\label{tab_photlog}
\end{deluxetable*}
%%%%%%%%%%%%%%%%%%%%%%%%%%%%%%%%%%%%%%%%%%%%%%%%%%%%%%%%%%%%%%%%%%%%%%%%%%%%%%%%

\begin{figure}
\centering
\includegraphics[width=0.48\textwidth]{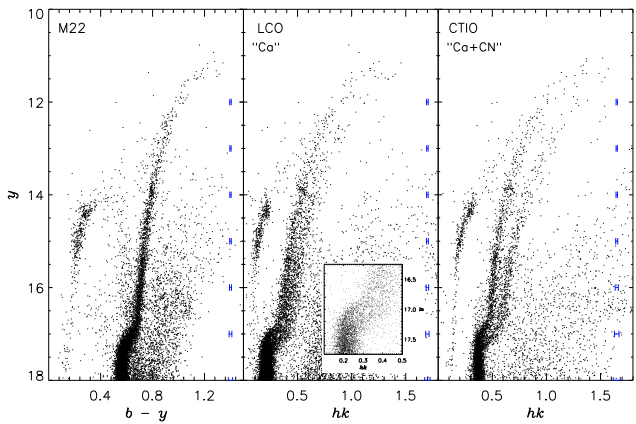}
\figcaption{CMDs for M22 in ($y$, $b-y$) and ($y$, $hk$) planes obtained with the Ca filter set at LCO is compared with ($y$, $hk$) CMD obtained with Ca+CN filter set at CTIO.
Note that the CMD obtained from the Ca+CN filter shows larger separation between the two RGBs than that obtained from the Ca filter.
The SGB stars are also split into two groups (see inset).
The horizontal bars denote the measurement error ($\pm$1$\sigma$).
\label{fig_m22cmd}}
\end{figure}

\begin{figure}
\centering
\includegraphics[width=0.48\textwidth]{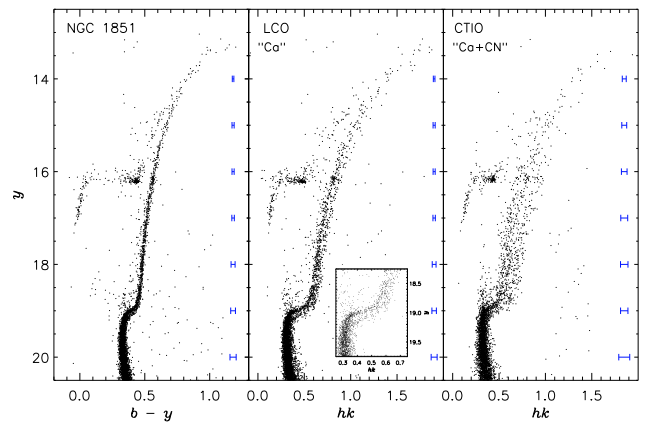}
\figcaption{Same as Figure~\ref{fig_m22cmd}, but for NGC 1851.
Note again that the separation between the two RGBs in the Ca+CN filter is larger than that in the Ca filter and that the bluer RGB stars in the Ca+CN filter show a broader distribution in the $hk$ index.
\label{fig_n1851cmd}}
\end{figure}

\begin{figure}
\centering
\includegraphics[width=0.48\textwidth]{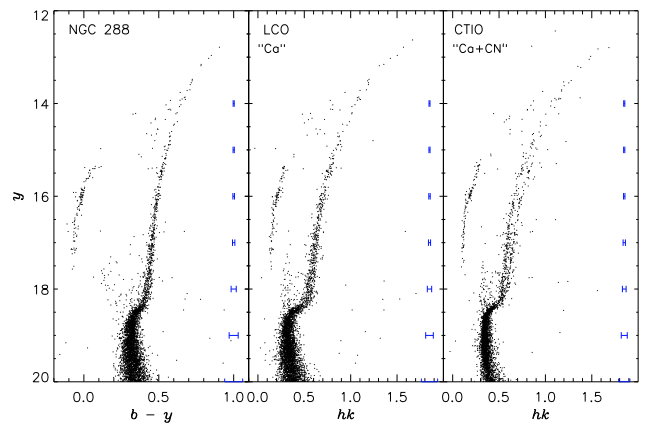}
\figcaption{Same as Figure~\ref{fig_m22cmd}, but for NGC 288.
Note that the RGB split reported earlier from the Ca+CN filter is not shown from the new Ca filter.
\label{fig_n288cmd}}
\end{figure}

Figures~\ref{fig_m22cmd}--\ref{fig_n288cmd} shows CMDs for M22, NGC 1851, and NGC 288 in ($y$, $b-y$) and ($y$, $hk$) planes obtained with the Ca and Ca+CN filter sets, respectively, where the $hk$ index, following \citet{Ant91}, is defined as $hk=($Ca$-b)-(b-y)$.
We used the separation index \citep{Ste03}, SHARP, and CHI parameters (i.e., $sep$ $<$ 1.0, $\chi$ $>$ 1.5, and SHARP $>$ $|0.25|$) to remove stars severely affected by adjacent starlight and non-stellar objects.
For M22 and NGC 1851, CMDs obtained from both Ca and Ca+CN filters show two distinct RGBs (see Figures~\ref{fig_m22cmd} and \ref{fig_n1851cmd}).
Note, however, that the separation between the two RGBs from the Ca filter is narrower than that from the Ca+CN filter (0.15 mag versus 0.20 mag for M22; 0.07 mag versus 0.16 mag for NGC 1851).
Since cross-matching of common stars between the two data sets indicates no change in the membership of the stars belonging to each subgroup (i.e., bluer and redder RGBs), this difference should be due to the difference in filter response function between the two filters.
We also detected the sub giant branch (SGB) split\footnote{It is important to note that the calcium abundance difference might have little to do with this split on the SGB. The origin of this split is suggested to be due to the C+N+O variation \citep{Yong09,Yong14a,Mar11b,Mar12,Mar14,Pio12}, age difference \citep{Mil08,Car10b,Gra12b}, and/or combined effects of metallicity, helium abundance, and [CNO/Fe] \citep{JL13}.}, reported earlier \citep{Mil08,Pio09,Mar09}, in our ($y$, $hk$) CMDs for M22 and NGC 1851 (Figures~\ref{fig_m22cmd} and \ref{fig_n1851cmd} insets).
The number ratios between the brighter and fainter SGB stars (0.59:0.41 for M22; 0.64:0.36 for NGC 1851)\footnote{This was done by simple eye count for the stars between 0.25 $<$ $hk$ $<$ 0.325 for M22 and 0.43 $<$ $hk$ $<$ 0.525 for NGC 1851. More rigorous analysis can be found in \citet{Mil09}.} and that between the bluer and redder RGB stars are roughly identical to the values reported in the literature \citep{Mil08,Mar09,Mar11b,Car11a,Gra14}.
Contrary to the cases of M22 and NGC 1851, the RGB split in NGC 288 is not evident in the new Ca photometry.
It appears, therefore, that the difference in calcium abundance, once suspected in our previous investigation \citep{Roh11}, would be negligible in NGC 288.
This suggests that the difference in CN abundance was mainly responsible for the RGB split reported in NGC 288 from the old Ca filter (Ca+CN filter) \citep[see also][]{Hsyu14}.
The decrease or disappearance of the separation between the two RGBs from the new Ca filter is consistent with the suspicion that the old photometry was severely affected by CN band.
In order to clarify this situation and also to investigate the origin of the RGB splits, we have carried out low-resolution spectroscopy described in the following section.

%%%%%%%%%%%%%%%%%%%%%%%%%%%%%%% Figure 2 'M22 CMD' %%%%%%%%%%%%%%%%%%%%%%%%%%%%%%%%%%%
%%%%%%%%%%%%%%%%%%%%%%%%%%%%%%%%%%%%%%%%%%%%%%%%%%%%%%%%%%%%%%%%%%%%%%%%%%%%%%%%

%%%%%%%%%%%%%%%%%%%%%%%%%%%%%%% Figure 3 'NGC 1851 CMD' %%%%%%%%%%%%%%%%%%%%%%%%%%%%%%%%%%%
%%%%%%%%%%%%%%%%%%%%%%%%%%%%%%%%%%%%%%%%%%%%%%%%%%%%%%%%%%%%%%%%%%%%%%%%%%%%%%%%

%%%%%%%%%%%%%%%%%%%%%%%%%%%%%%% Figure 4 'NGC 288 CMD' %%%%%%%%%%%%%%%%%%%%%%%%%%%%%%%%%%%
%%%%%%%%%%%%%%%%%%%%%%%%%%%%%%%%%%%%%%%%%%%%%%%%%%%%%%%%%%%%%%%%%%%%%%%%%%%%%%%%

% Section 3 (Low-Resolution Spectroscopy) 
\section{Low-Resolution Spectroscopy}
\label{sec_spectroscopy}

%%%%%%%%%%%%%%%%%%%%%%%%%%%%%%% Figure 5 'target CMD' %%%%%%%%%%%%%%%%%%%%%%%%%%%%%%%%%%%
\begin{figure*}
\centering
\includegraphics[width=0.95\textwidth]{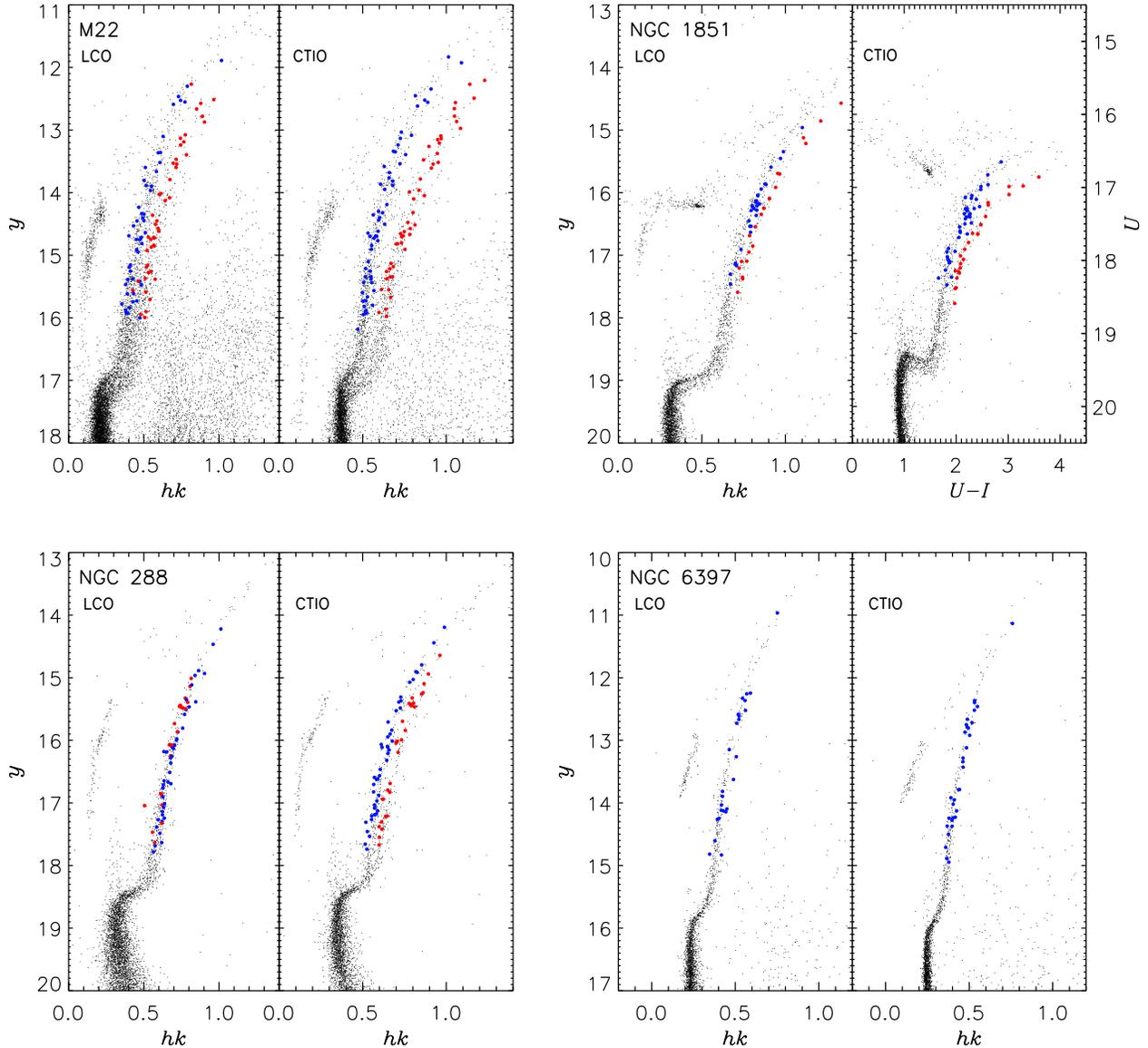}
\figcaption{Spectroscopic target stars identified on the CMDs for M22, NGC 1851, NGC 288, and NGC 6397.
Photometry obtained from LCO is compared with that from CTIO.
Sample selection is based on the CTIO photometry (see the text).
Blue and red circles indicate selected stars from the bluer and redder RGBs, respectively.
\label{fig_target_cmd}}
\end{figure*}
%%%%%%%%%%%%%%%%%%%%%%%%%%%%%%%%%%%%%%%%%%%%%%%%%%%%%%%%%%%%%%%%%%%%%%%%%%%%%%%%

% Section 3.1 (Observations and Data Reduction)
\subsection{Observations and Data Reduction}
\label{sec_spectroscopy_obs}
The spectral data were taken from multi-object spectroscopic mode with Wide Field Reimaging CCD Camera (WFCCD) mounted on the du Pont telescope.
Figure~\ref{fig_target_cmd} shows our spectroscopic target stars on the CMDs for M22, NGC 1851, NGC 288, and NGC 6397.
Target stars were selected from two distinct subpopulations in the ($y$, $hk$) CMDs using Ca+CN filter, for M22 and NGC 288, and in the ($U$, $U-I$) CMD, for NGC 1851, all obtained from CTIO 4m telescope.
These filter combinations provide the most clear separation between the two RGBs, and as shown in Figure~\ref{fig_target_cmd}, these sub groupings are almost identical to those obtained from new Ca filter.
In our selection of target stars, a similar number of stars were selected from the two subpopulations.
Four or five slit masks, each of which was made for stars with similar brightness, were designed using the $maskgenx4$ code.
In addition, as a comparison, we made two slit masks for NGC 6397, which shows only a single narrow RGB in all photometry.
Typically, each mask includes $\sim$25 slits of 1.2{\arcsec} width with slit length longer than 10{\arcsec} for obtaining sky spectrum.
More than 5 slit boxes of 12{\arcsec} $\times$ 12{\arcsec} were also made for alignment stars.
The name of mask and number of containing stars are listed in Table~\ref{tab_speclog}.

%%%%%%%%%%%%%%%%%%%%%%%%%%%%%%% Table 'log.spectroscopy' %%%%%%%%%%%%%%%%%%%%%%%%%%%%%%%%%%%%
\begin{deluxetable}{ccccc}
\tabletypesize{\small}
%\tablewidth{0pt}
\tablecaption{Mask Descriptions and Spectroscopic Observation Log}
\tablehead{
\colhead{Object} & \colhead{Mask} & \colhead{\#stars} & \colhead{Exposures (N$\times$s)}
}
\startdata
M22      & Bright    & 24 & 6$\times$1200 \\
         & Bright-II & 18 & 5$\times$1200 \\
         & Faint     & 27 & 5$\times$1200 \\
         & Faint-II  & 21 & 6$\times$1200 \\
         & Faint-III & 25 & 6$\times$1200 \\
NGC 288  & Bright    & 23 & 4$\times$1200 \\
         & Bright-II & 19 & 5$\times$1200 \\
         & Faint     & 23 & 4$\times$1500 \\
         & Faint-II  & 27 & 3$\times$1800 \\
NGC 1851 & Bright    & 24 & 5$\times$1200 \\
         & Bright-II & 18 & 3$\times$1500 \\
    & Intermediate   & 24 & 7$\times$1500 \\
         & Faint     & 30 & 3$\times$1800
\enddata
\label{tab_speclog}
\end{deluxetable}
%%%%%%%%%%%%%%%%%%%%%%%%%%%%%%%%%%%%%%%%%%%%%%%%%%%%%%%%%%%%%%%%%%%%%%%%%%%%%%%%

%%%%%%%%%%%%%%%%%%%%%%%%%%%%%%% Figure 6 'Index Definition' %%%%%%%%%%%%%%%%%%%%%%%%%%%%%%%%%%%
\begin{figure}
\centering
\includegraphics[width=0.48\textwidth]{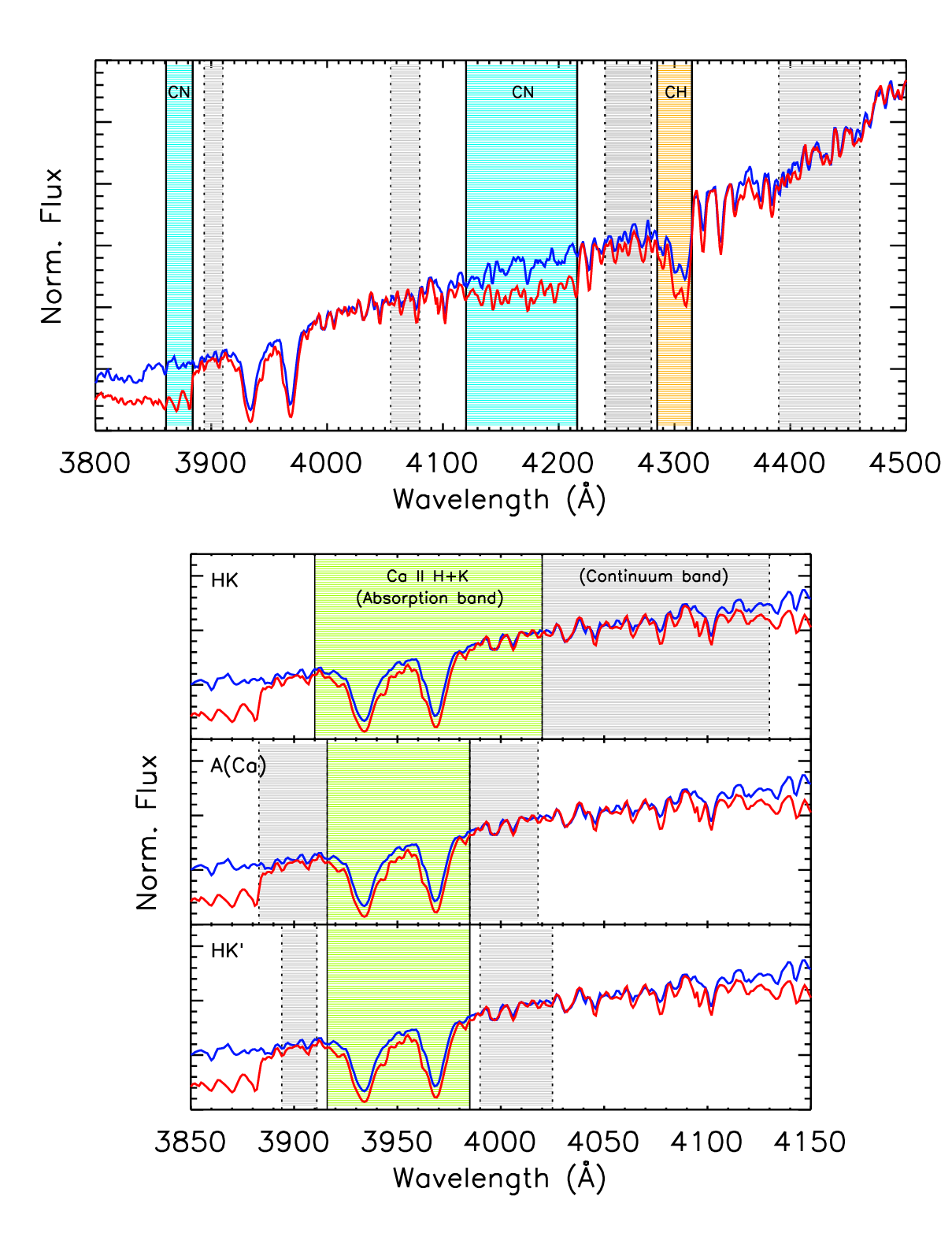}
\figcaption{Upper panel: The absorption bands for CN (cyan) and CH (red-yellow) indices, together with their continuum bands (grey) are shown with two example spectra.
Two spectra in blue and red are from two stars, at similar magnitudes, in bluer and redder RGBs in M22, respectively.
Lower panels: The comparison of three calcium indices, HK (top), A(Ca) (middle), and HK$'$ (bottom).
In each panel, absorption and continuum bands are represented in green and grey, respectively.
In the present study, we have used the HK$'$ index (see the text).
\label{fig_index}}
\end{figure}
%%%%%%%%%%%%%%%%%%%%%%%%%%%%%%%%%%%%%%%%%%%%%%%%%%%%%%%%%%%%%%%%%%%%%%%%%%%%%%%%

The observations were carried out during four observing runs of 2011 July 7--13, 2011 September 16--19, 2012 February 19--22, and 2013 June 15--21.
The detector was a 4064 $\times$ 4064 WF4K CCD providing a plate scale of 0.484{\arcsec}/pixel and a FOV of 25{\arcmin} $\times$ 25{\arcmin}.
We used the H\&K grism with a dispersion of 0.8 {\AA}/pixel and a central wavelength of 3700 {\AA}.
The spectral coverage depends on the location of a slit on the mask, but the calcium II H \& K lines, CN and CH bands (3700 {\AA} $\sim$ 4500 {\AA}) are covered for all sample stars.
The exposure times varied from 1200 to 1800 seconds, depending on the brightness grouping of stars, and at least three exposures were taken for each mask (see Table~\ref{tab_speclog}).
In addition, calibration frames, three flats and an arc lamp, were obtained for each mask before/after exposures.

The data reduction was performed with IRAF and the modified WFCCD reduction package, for which the detailed description, including flat fielding, distortion correction, wavelength calibration, and sky subtraction, can be found in \citet{Pro06}.
Wavelength solution was calculated from the emission lines of He+Ne arc lamp, and cosmic rays were removed by using the $lacosmic$ task \citep{Van01}.
Spectral energy distributions were obtained after subtracting sky spectrum, which was taken from two sky areas far from the target star in the same slit.
Note that flux calibration and continuum normalization were not necessary for our analysis, because spectral indices in our low-resolution spectroscopy are defined as the ratio of absorption strength to nearby continuum \citep{Har03,Kay08,Pan10}.
Radial velocities were measured from several strong absorption lines in our wavelength coverage, such as calcium II H \& K, and H$\beta$ lines, by using the $rvidlines$ task in IRAF RV package.
These radial velocities were also used to remove non-member stars from the 2.5 sigma-clipping rejection.
Signal-to-noise (S/N) ratio was measured at  $\sim$3900 {\AA}, and low S/N stars (S/N ratio $<$ 8) were excluded in our analysis.

%%%%%%%%%%%%%%%%%%%%%%%%%%%%%%% Figure 7 'HK'-[Fe/H]' %%%%%%%%%%%%%%%%%%%%%%%%%%%%%%%%%%%
\begin{figure}
\centering
\includegraphics[width=0.48\textwidth]{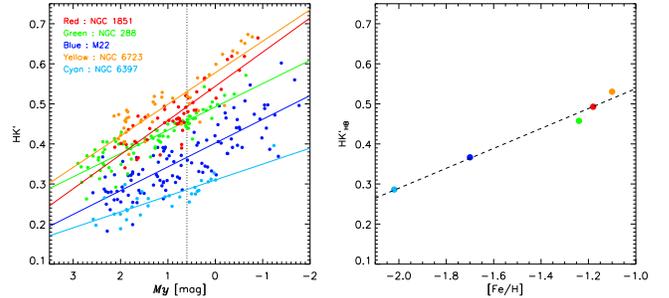}
\figcaption{Left panel: The measured HK$'$ index as a function of absolute $y$ magnitudes is presented for five sample GCs.
Solid lines are least square fits and the vertical dotted line is the approximate magnitude level of horizontal branch (HB).
Distance moduli were adopted from the updated catalog of \citet{Har96}.
Right panel: The HK$'$ index at the HB level is plotted as a function of [Fe/H] \citep[adopted from][]{Har96}.
The dotted line is a least square fit (HK$'$=0.79+0.25[Fe/H]).
\label{fig_feh_hk}}
\end{figure}
%%%%%%%%%%%%%%%%%%%%%%%%%%%%%%%%%%%%%%%%%%%%%%%%%%%%%%%%%%%%%%%%%%%%%%%%%%%%%%%%

%%%%%%%%%%%%%%%%%%%%%%%%%%%%%%% Table 'all_index' %%%%%%%%%%%%%%%%%%%%%%%%%%%%%%%%%%%%
\begin{deluxetable*}{cccccccccccccc}
\tabletypesize{\scriptsize}
%\tablewidth{0pt}
\tablecaption{Index Measurements for the sample stars in M22 and NGC 288}
\tablehead{
\colhead{ID} & \colhead{Ra} & \colhead{Dec} & \colhead{$y$} & \colhead{$hk$} & \colhead{HK$'$} & \colhead{errHK$'$} & \colhead{$\delta$HK$'$} & \colhead{CN} & \colhead{errCN} & \colhead{$\delta$CN} & \colhead{CH} & \colhead{errCH} & \colhead{$\delta$CH}% & \colhead{Pop}
}
\startdata
%### M22 ###
M22-1001 & 279.04150 & -23.88145 & 11.93 & 1.091 &  0.5600 &  0.0066 &  0.0584 & -0.1762 &  0.0108 & -0.3839 &  1.1380 &  0.0046 & -0.0366 \\
M22-1009 & 279.05707 & -23.97182 & 13.08 & 0.792 &  0.4371 &  0.0096 &  0.0042 & -0.0735 &  0.0135 & -0.1540 &  1.0390 &  0.0066 & -0.0807 \\
M22-1025 & 279.07031 & -23.87054 & 12.62 & 0.828 &  0.4150 &  0.0067 & -0.0455 & -0.1409 &  0.0098 & -0.2726 &  1.0427 &  0.0046 & -0.0991 \\
M22-2001 & 279.04425 & -23.94319 & 13.09 & 0.967 &  0.4415 &  0.0117 &  0.0093 &  0.2320 &  0.0134 &  0.1527 &  1.2347 &  0.0072 &  0.1155 \\
M22-2002 & 279.04752 & -23.92560 & 13.54 & 0.922 &  0.4481 &  0.0105 &  0.0426 &  0.3652 &  0.0111 &  0.3354 &  1.1556 &  0.0068 &  0.0577 \\
M22-2003 & 279.05527 & -23.92344 & 13.48 & 0.864 &  0.4021 &  0.0111 & -0.0074 &  0.1730 &  0.0129 &  0.1358 &  1.1686 &  0.0069 &  0.0676 \\
...      & ...       &  ...      & ...   & ...   &  ...    &  ...    &  ...    &  ...    &  ...    &  ...    &  ...    &  ...    &  ...    \\
%### NGC 288 ###
N288-1001 &  13.20749 & -26.61549 & 14.19 & 0.989 &  0.5713 &  0.0125 &  0.0413 & -0.0326 &  0.0187 & -0.0529 &  1.2075 &  0.0085 &  0.0467 \\
N288-1003 &  13.22497 & -26.57806 & 14.44 & 0.926 &  0.5245 &  0.0139 &  0.0089 & -0.1092 &  0.0213 & -0.1062 &  1.1466 &  0.0096 & -0.0045 \\
N288-1006 &  13.14758 & -26.62872 & 14.80 & 0.853 &  0.4983 &  0.0155 &  0.0032 & -0.1993 &  0.0248 & -0.1632 &  1.1506 &  0.0104 &  0.0132 \\
N288-2001 &  13.30425 & -26.61799 & 14.64 & 0.962 &  0.5212 &  0.0144 &  0.0173 &  0.2212 &  0.0175 &  0.2430 &  1.0789 &  0.0102 & -0.0645 \\
N288-2002 &  13.16385 & -26.63292 & 14.94 & 0.893 &  0.4927 &  0.0156 &  0.0061 &  0.1788 &  0.0192 &  0.2285 &  1.0886 &  0.0108 & -0.0432 \\
N288-2003 &  13.27368 & -26.57371 & 15.10 & 0.867 &  0.4858 &  0.0163 &  0.0083 &  0.2131 &  0.0195 &  0.2773 &  1.0930 &  0.0112 & -0.0327 \\
...       & ...       &  ...      & ...   & ...   &  ...    &  ...    &  ...    &  ...    &  ...    &  ...    &  ...    &  ...    &  ...    
\enddata
\label{tab_index_all}
\end{deluxetable*}
%%%%%%%%%%%%%%%%%%%%%%%%%%%%%%%%%%%%%%%%%%%%%%%%%%%%%%%%%%%%%%%%%%%%%%%%%%%%%%%%

%%%%%%%%%%%%%%%%%%%%%%%%%%%%%%% Table '1851_index' %%%%%%%%%%%%%%%%%%%%%%%%%%%%%%%%%%%%
\begin{deluxetable*}{cccccccccccccc}
\setlength{\tabcolsep}{0.04in}
\tabletypesize{\scriptsize}
\tablewidth{0pt}
\tablecaption{Index Measurements for the sample stars in NGC 1851}
\tablehead{
\colhead{ID} & \colhead{Ra} & \colhead{Dec} & \colhead{$V$} & \colhead{$U$-$I$} & \colhead{HK$'$} & \colhead{errHK$'$} & \colhead{$\delta$HK$'$} & \colhead{CN} & \colhead{errCN} & \colhead{$\delta$CN} & \colhead{CH} & \colhead{errCH} & \colhead{$\delta$CH}% & \colhead{Pop}
}
\startdata
%### NGC 1851 ###
N1851-1002 &  78.43302 & -40.02103 & 15.96 & 2.346 &  0.5148 &  0.0205 &  0.0166 &  0.1137 &  0.0267 &  0.0189 &  1.0230 &  0.0150 & -0.0718 \\
N1851-1003 &  78.45161 & -40.05419 & 15.42 & 2.613 &  0.5143 &  0.0194 & -0.0270 &  0.0708 &  0.0260 & -0.1026 &  1.1362 &  0.0133 &  0.0174 \\
N1851-1005 &  78.46117 & -40.11313 & 16.16 & 2.264 &  0.4570 &  0.0214 & -0.0252 &  0.0240 &  0.0286 & -0.0417 &  1.0800 &  0.0146 & -0.0059 \\
N1851-2001 &  78.44778 & -40.06932 & 16.04 & 2.474 &  0.5204 &  0.0215 &  0.0291 &  0.3157 &  0.0245 &  0.2334 &  1.1026 &  0.0151 &  0.0116 \\
N1851-2002 &  78.45982 & -40.03528 & 14.82 & 3.289 &  0.6185 &  0.0193 &  0.0284 &  0.6024 &  0.0196 &  0.3399 &  1.2042 &  0.0136 &  0.0583 \\
N1851-2005 &  78.47660 & -40.03832 & 15.66 & 2.617 &  0.5615 &  0.0205 &  0.0396 &  0.3878 &  0.0229 &  0.2497 &  1.0795 &  0.0149 & -0.0285 \\
...        & ...       &  ...      & ...   & ...   &  ...    &  ...    &  ...    &  ...    &  ...    &  ...    &  ...    &  ...    &  ...    
\enddata
\label{tab_index_1851}
\end{deluxetable*}
%%%%%%%%%%%%%%%%%%%%%%%%%%%%%%%%%%%%%%%%%%%%%%%%%%%%%%%%%%%%%%%%%%%%%%%%%%%%%%%%

% Section 3.2 (Definition of Spectral Indices)
\subsection{Definition of Spectral Indices}
\label{sec_spectroscopy_index}
Since the calcium II H \& K lines are one of the strongest absorption lines in stellar spectrum, several index definitions are available in the literature for these lines.
The most widely used indices are HK index defined by \citet{Sun80}, and A(Ca) index defined by \citet{NF82}.
However, we found that the wavelength range of absorption band adapted in HK index is too broad to detect a subtle difference in calcium abundance, whereas A(Ca) index can be more influenced by CN band as the blue side of the continuum is too close to that band.
Therefore, we defined a new calcium index (HK$'$), which is analogous to A(Ca) index except the width of blue side continuum band.
The definition of this index is
\begin{eqnarray*} 
{\rm HK'}  & = & -2.5 \log{\frac{F_{3916-3985}}{2F_{3894-3911}+F_{3990-4025}}} ,
\end{eqnarray*}
where $F_{3916-3985}$, for example, is the integrated flux from 3916 to 3985 {\AA}.
In lower panel of Figure~\ref{fig_index}, our new index is compared with other indices.
The final HK$'$ index for target star was obtained by taking error-weighted mean of indices measured from each exposure, where the measurement error was estimated assuming Poisson statistics in the flux measurements \citep{VE06}.

As well known, spectral indices are affected not only by chemical abundance but also by effective temperature ($T_{\rm eff}$) and surface gravity ($\log g$).
Therefore, in order to measure the index difference only from the abundance effect, we used $\delta$HK$'$ index, at fixed temperature and gravity, as employed by previous studies \citep{Nor81,NF83,Har03}.
The $\delta$HK$'$ index was calculated as the difference between HK$'$ index for each star and the least-square fitting line, obtained from the full sample in a GC on the HK$'$ versus magnitude diagram.
Furthermore, we derived the relation, HK$'$$=$0.79$+$0.25[Fe/H], to convert our result from HK$'$ index to [Fe/H] from our sample GCs (see Figure~\ref{fig_feh_hk}).

Similarly to the calcium index, several different definitions of CN and CH indices are given in the literature \citep{NF79,Nor81,Coh99a,Har03}.
In this study, we used the most widely used indices, S3839 and S4142 for CN band strength, and CH(4300) for CH band, as defined by \citet{Har03}.
The definitions for these indices are
\begin{eqnarray*}
{\rm S}(3839) & = & -2.5 \log{\frac{F_{3861-3884}}{F_{3894-3910}}} , \\
{\rm S}(4142) & = & -2.5 \log{\frac{F_{4120-4216}}{0.4F_{4055-4080}+0.6F_{4240-4280}}} , \\
{\rm CH4300} & = & -2.5 \log{\frac{F_{4285-4315}}{0.5F_{4240-4280}+0.5F_{4390-4460}}} .
\end{eqnarray*}
In general, two CN indices show tight correlation, however, previous studies \citep{Har03,Pan10,Lar12} have shown that the S3839 index shows higher sensitivity and smaller error than S4142 index.
Therefore, we have used S3839 as a main CN index.
Absorption bands adapted in our CN and CH indices are illustrated in Figure~\ref{fig_index} (upper panel), and the measured indices and errors are listed in Tables~\ref{tab_index_all} and \ref{tab_index_1851}.

% Section 4 (Differences in HK’ and CN indices between the two subpopulations)
\section{Differences in HK$'$ and CN indices between the two subpopulations}
\label{sec_result}

% Section 4.1 (M22)
\subsection{M22}
\label{sec_result_m22}

%%%%%%%%%%%%%%%%%%%%%%%%%%%%%%% Figure 8 'M22 Spectroscopy' %%%%%%%%%%%%%%%%%%%%%%%%%%%%%%%%%%%
\begin{figure}
\centering
\includegraphics[width=0.48\textwidth]{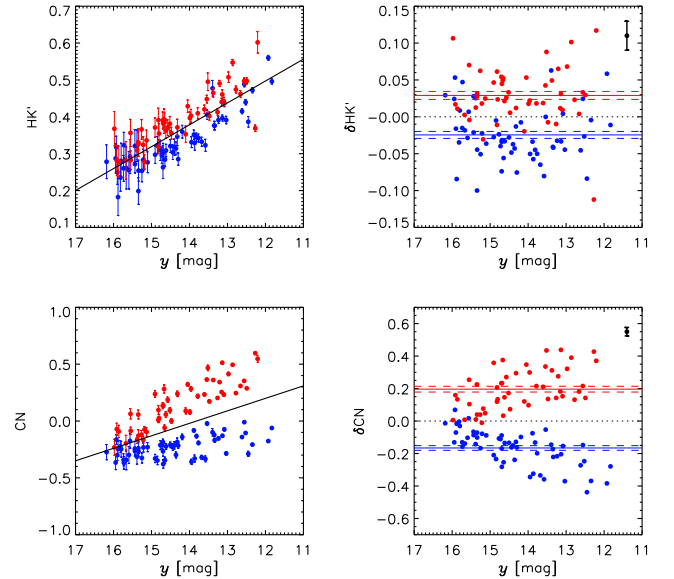}
\figcaption{Measured spectral indices (HK$'$ and CN) as functions of $y$ magnitude for M22.
Upper panels: The $\delta$HK$'$ indices plotted against $y$ magnitude, where the blue and red circles are stars in bluer and redder RGBs in Figure~\ref{fig_target_cmd}.
The $\delta$HK$'$ index is defined as the height of the HK$'$ index above the least square line in the left panel.
The mean value and the error of the mean ($\pm$1$\sigma$) for each subpopulation are denoted by the solid and dashed lines, respectively.
The vertical bar in the right panel denotes the typical measurement error.
Lower panels: Same as the upper panels, but for the CN (left) and $\delta$CN (right) indices.
Note that the two subpopulations are separated in $\delta$HK$'$ and $\delta$CN indices, at 7.5$\sigma$ and 15.9$\sigma$ levels, respectively.
\label{fig_m22spec}}
\end{figure}
%%%%%%%%%%%%%%%%%%%%%%%%%%%%%%%%%%%%%%%%%%%%%%%%%%%%%%%%%%%%%%%%%%%%%%%%%%%%%%%%

Early spectroscopic observations indicated that M22 shows abundance variations in light elements from CN and CH bands \citep{Coh81,Pil82,NF83,Leh91,Ant95}.
Some spreads in the abundance of heavy elements, such as calcium and iron, have also been suspected from these studies, and this is confirmed from recent high and intermediate resolution spectroscopy \citep{Mar09,Mar11b,Da09}.
The narrow-band Ca photometry has reported a clear split between the two subpopulations \citep{jwlee09a}, but it was not clear whether this was solely due to the difference in calcium abundance because of the contamination from CN band (see Section~\ref{sec_photometry}).
Our new Ca photometry, reported in Section~\ref{sec_photometry}, has shown that this split is maintained, although the difference in the $hk$ index is reduced between the two RGBs.
In order to investigate the origin of this split, we have compared spectral indices between the two subpopulations.

Figure~\ref{fig_m22spec} shows measured HK$'$, $\delta$HK$'$, CN, and $\delta$CN indices for the RGB stars in M22 as functions of $y$ magnitude.
Note again that even for the RGB stars having same chemical composition, HK$'$ and CN indices would increase with decreasing magnitude because of the temperature effect (see Section~\ref{sec_spectroscopy_index}).
Therefore, the differences in chemical abundance between the two subpopulations are measured on the $\delta$HK$'$ and $\delta$CN indices versus magnitude diagrams by taking the mean values for each subpopulation.
The upper right panel of Figure~\ref{fig_m22spec} clearly shows that the two subpopulations are separated by 0.054 in $\delta$HK$'$ index, which is significant at 7.5$\sigma$ level.
The fact that stars on the redder RGB show higher $\delta$HK$'$ values than those on the bluer RGB suggests that the separation found in our Ca photometry is indeed originated from the difference in calcium abundance.
In terms of $\Delta$[Fe/H], this difference in $\delta$HK$'$ index is equivalent to 0.18 dex (see Section~\ref{sec_spectroscopy_index}).
This is similar to the results obtained from high and intermediate resolution spectroscopy (0.15 $\sim$ 0.26 dex; \citealt{Mar09,Mar11b,Da09}).
Ca-strong stars in our sample also show stronger CN band strength (see the lower right panel of Figure~\ref{fig_m22spec}).
We can also see a clear separation between the two subpopulations in $\delta$CN index ($\sim$0.36), which is significant at 15.9$\sigma$ level.
This result, based on a lager sample, therefore confirms the presence of CN bimodality reported earlier in this cluster \citep{NF83,Kay08}.

%%%%%%%%%%%%%%%%%%%%%%%%%%%%%%% Figure 9 'NGC 1851 Spectroscopy' %%%%%%%%%%%%%%%%%%%%%%%%%%%%%%%%%%%
\begin{figure}
\centering
\includegraphics[width=0.48\textwidth]{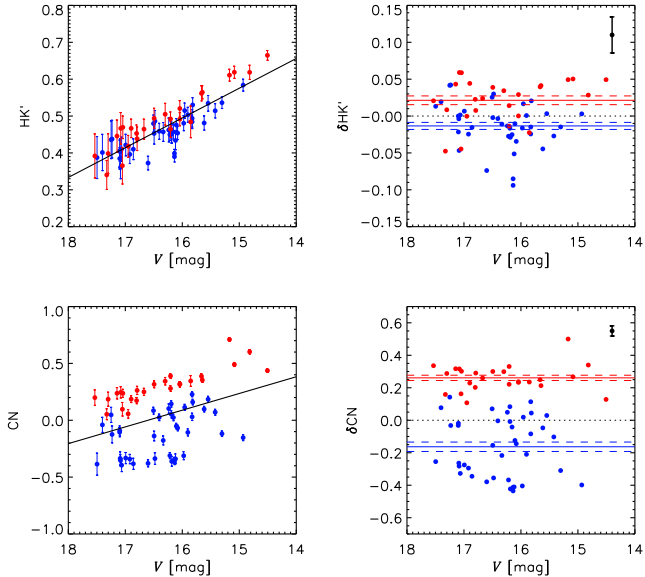}
\figcaption{Same as Figure~\ref{fig_m22spec}, but for NGC 1851.
Similar to the case of M22, the two subpopulations are separated in $\delta$HK$'$ and $\delta$CN indices, at 4.6$\sigma$ and 12.8$\sigma$ levels, respectively.
\label{fig_n1851spec}}
\end{figure}
%%%%%%%%%%%%%%%%%%%%%%%%%%%%%%%%%%%%%%%%%%%%%%%%%%%%%%%%%%%%%%%%%%%%%%%%%%%%%%%%

%%%%%%%%%%%%%%%%%%%%%%%%%%%%%%% Figure 10 'N1851 CN-histogram' %%%%%%%%%%%%%%%%%%%%%%%%%%%%%%%%%%%
\begin{figure}
\centering
\includegraphics[width=0.45\textwidth]{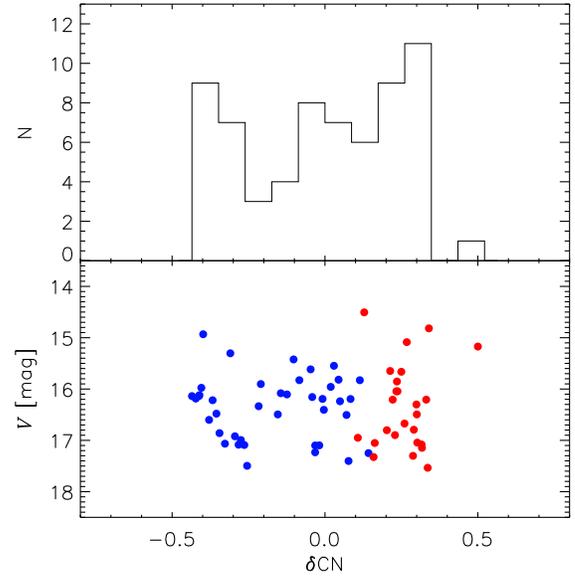}
\figcaption{The distribution of $\delta$CN index against $V$ magnitude for NGC 1851.
The histogram in the upper panel suggests the presence of three subpopulations.
\label{fig_n1851hist}}
\end{figure}
%%%%%%%%%%%%%%%%%%%%%%%%%%%%%%%%%%%%%%%%%%%%%%%%%%%%%%%%%%%%%%%%%%%%%%%%%%%%%%%%

% Section 4.2 (NGC 1851)
\subsection{NGC 1851}
\label{sec_result_N1851}
Since the early detection of CN-strong stars in NGC 1851 \citep{Hes82}, recent spectroscopic observations found abundance variations in many light elements from MS to RGB stars in this GC \citep{YG08,Yong09,Pan10,Vil10,Camp12,Lar12}.
In addition, \citet{Car10b,Car11b} detected the spread in [Fe/H] among RGB stars.
Several photometric observations, including our Ca photometry reported in Section~\ref{sec_photometry}, also show a clear split in RGB \citep{Cal07,Han09,jwlee09b,Car11a}.
In order to understand the origin of this split, we have again compared spectral indices between the two subpopulations.

Figure~\ref{fig_n1851spec} shows distributions of measured spectral indices for stars on the two RGBs in NGC 1851.
Similar to the case of M22, these stars are separated in both $\delta$HK$'$ and $\delta$CN indices.
The difference between the two subpopulations in $\delta$HK$'$ index is 0.035 and that in $\delta$CN index is 0.42, which are significant at 4.6$\sigma$ and 12.8$\sigma$ levels, respectively.
We also estimated the difference in [Fe/H], $\sim$0.14 dex, from the difference in $\delta$HK$'$ index, which is somewhat larger than the difference obtained from high-resolution spectroscopy (0.06 $\sim$ 0.08 dex; \citealt{Car10b}).
Our results for this GC suggest that stars on the two distinct RGBs, divided in both the $U-I$ color and the $hk$ index, indeed have different calcium abundances.
The RGB split in the $U-I$ color is most likely affected by the difference in CN band strength as the wide passband of the $U$ filter includes this molecular band, while the split in the $hk$ index is originated from the difference in calcium abundance.
In addition, Figure~\ref{fig_n1851hist} shows the histogram for $\delta$CN index, which suggests that RGB stars in NGC 1851 would be divided into three subpopulations.
This is mainly because the bluer RGB stars in our Ca photometry are further divided into two subpopulations.
This trimodal distribution is comparable with the result by \citet{Camp12}.
The presence of three subpopulations, instead of two, in massive GCs is also suggested from the recent population models \citep{Jang14}, and the observed Na-O anti-correlation \citep{Car09b}.

%%%%%%%%%%%%%%%%%%%%%%%%%%%%%%% Figure 11 'NGC 288 Spectroscopy' %%%%%%%%%%%%%%%%%%%%%%%%%%%%%%%%%%%
\begin{figure}
\centering
\includegraphics[width=0.48\textwidth]{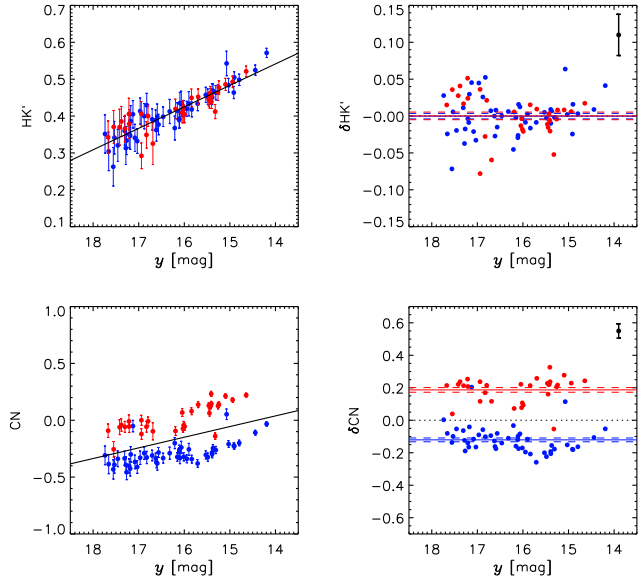}
\figcaption{Same as Figure~\ref{fig_m22spec}, but for NGC 288.
Note that the two subpopulations show a clear separation only in the $\delta$CN index (16.6$\sigma$).
\label{fig_n288spec}}
\end{figure}
%%%%%%%%%%%%%%%%%%%%%%%%%%%%%%%%%%%%%%%%%%%%%%%%%%%%%%%%%%%%%%%%%%%%%%%%%%%%%%%%

% Section 4.3 (NGC 288 & NGC 6397)
\subsection{NGC 288 \& NGC 6397}
\label{sec_result_n288}
As described in Section~\ref{sec_photometry}, the RGB split in NGC 288, discovered from the old Ca photometry \citep{Roh11}, is not confirmed from the new Ca photometry reported in this study.
Figure~\ref{fig_n288spec} shows a very clear separation in $\delta$CN index among RGB stars in this GC, while this is not shown in the $\delta$HK$'$ index.
The difference in $\delta$CN index between the two subpopulations is 0.31, which is significant at 16.6$\sigma$ level.
This result is comparable to the spread in $\delta$CN reported by \citet{Kay08} and \citet{SL09}.
The mean $\delta$HK$'$ indices of the two subpopulations, however, are almost identical well within the standard error, suggesting that the difference in $\delta$HK$'$ index is negligible.
This is consistent with the recent result from high-resolution spectroscopy report by \citet{Hsyu14}.
Therefore, unlike M22 and NGC 1851, it appears that there is no evidence for the difference in calcium abundance in this GC.
This result, together with our new Ca photometry, indicates that the origin of the RGB split reported in old photometry was solely due to the effect of CN band.
At the same time, in other respects, this also verifies that our new photometry is not contaminated by CN band.

\begin{figure}
\centering
\includegraphics[width=0.48\textwidth]{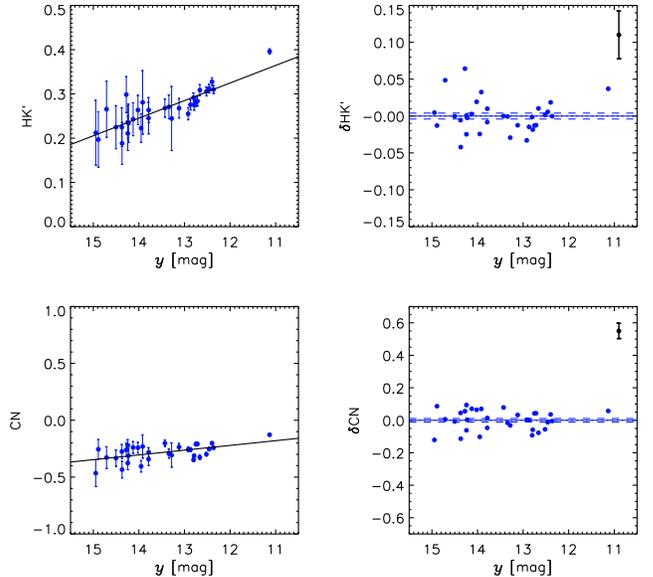}
\figcaption{Same as Figure~\ref{fig_m22spec}, but for NGC 6397.
There are no apparent spreads in $\delta$HK$'$ and $\delta$CN indices in this GC (see the text).
\label{fig_n6397spec}}
\end{figure}

For NGC 6397, which has only single RGB, and therefore stellar population is expected to be more homogeneous, our sample stars indeed show single and narrow distributions in all spectral indices (see Figure~\ref{fig_n6397spec}).
The standard deviation of the distribution in $\delta$HK$'$ is 0.023, which is similar to that for NGC 288 but smaller than those for M22 and NGC 1851 ($\sim$0.04).
In the case of $\delta$CN index, standard deviation for this GC ($\sim$0.062) is also much smaller than those for other GCs ($\sim$0.18).

%%%%%%%%%%%%%%%%%%%%%%%%%%%%%%% Figure 12 'NGC 6397 Spectroscopy' %%%%%%%%%%%%%%%%%%%%%%%%%%%%%%%%%%%
%%%%%%%%%%%%%%%%%%%%%%%%%%%%%%%%%%%%%%%%%%%%%%%%%%%%%%%%%%%%%%%%%%%%%%%%%%%%%%%%

%%%%%%%%%%%%%%%%%%%%%%%%%%%%%%% Figure 13 'CN-CH' %%%%%%%%%%%%%%%%%%%%%%%%%%%%%%%%%%%
\begin{figure*}
\centering
\includegraphics[width=0.95\textwidth]{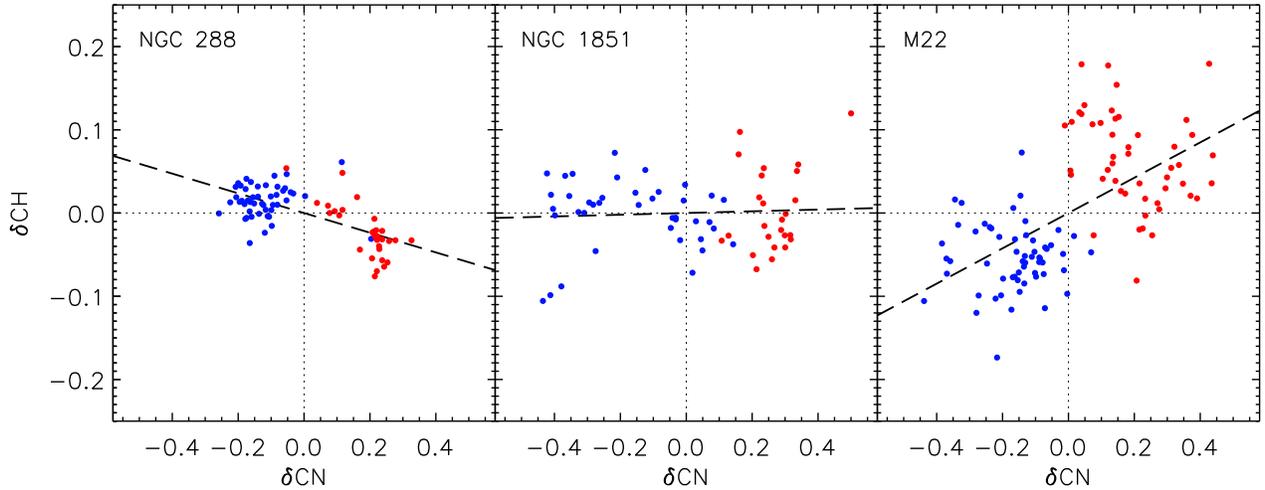}
\figcaption{The systematic difference in the CN-CH correlation among RGB stars in NGC 288, NGC 1851, and M22.
The RGB stars in NGC 288 show anti-correlation between the CN and CH bands, while those in M22 show positive correlation.
No apparent correlation is shown in NGC 1851.
In each panel, a dashed line represents the least-square fitting line, obtained from the full sample.
Symbol definitions are as in Figure~\ref{fig_target_cmd}.
\label{fig_cn_ch}}
\end{figure*}
%%%%%%%%%%%%%%%%%%%%%%%%%%%%%%%%%%%%%%%%%%%%%%%%%%%%%%%%%%%%%%%%%%%%%%%%%%%%%%%%

% Section 5 (CN-CH correlations and Anti-correlations)
\section{CN-CH Correlations and Anti-correlations}
\label{sec_cn_ch}
The CN-CH anti-correlation is one of the most well-known features in GC spectroscopic studies.
This feature was reported for many GCs, including M2, M3, M5, M13, M71, NGC 5927, NGC 6352, NGC 6752, Pal 12, and 47 Tuc \citep[][and references therein]{SS91,Smi96,Can98,Coh99a,Coh99b,Har03,Pan10,Smo11}.
Since CN band strength is mostly affected by nitrogen abundance while CH band is a measure of carbon abundance, the CN-CH anti-correlation would indicate that nitrogen abundance is inversely correlated with carbon abundance \citep{Smi96}.
This chemical disproportion is generally interpreted as a result of evolution mixing \citep{SM79,Cha95,DV03,jwlee10} or self-enrichment \citep{CD81,Can98}.
While the evolution mixing contributes to the presence of CN-CH anti-correltaion in RGB stars, recent studies favored the self-enrichment scenario over evolution mixing, because this feature is observed even from the un-evolved stage, MS and SGB \citep{Kay08,Pan10}, which suggests a primordial difference in chemical composition.
Also, the presence of distinct separation between the two groups would be hardly reproduced by the evolution scenario, as this mechanism is more likely to produce just the spread rather than the discrete distribution.

In order to investigate the correlation between CN and CH band strengths in our sample GCs, in Figure~\ref{fig_cn_ch}, we plot $\delta$CN versus $\delta$CH of RGB stars in NGC 288, NGC 1851, and M22, respectively.
In the case of NGC 288, we can see the anti-correlation between CN and CH \citep[see also][]{Kay08,SL09}, which is consistent with the general trend reported in previous studies.
NGC 1851, however, shows no apparent relation between CN and CH indices, mostly because the difference in $\delta$CH index is negligible between the two subpopulations.
This result is consistent with earlier findings from MS and SGB stars in this GC \citep{Pan10,Lar12}.
On the other hand, M22 apparently shows a positive correlation between CN and CH.
The correlation similar to this was reported by \citet{NF83} and \citet{Ant95} from their sample of RGB stars.
This was not detected, however, by \citet{Kay08} and \citet{Pan10} from their sample of MS and SGB stars, although the insufficient S/N ratio and the effect of differential reddening have prevented them to reach a firm conclusion.
Careful examination of the redder RGB subpopulation (red circle) in Figure~\ref{fig_cn_ch} indicates, however, a sign of anti-correlation between CN and CH.
This suggests that the apparent correlation has probably arisen because of the mixture of the two distinct subpopulations having different metal abundance (see below), each of which has CN-CH anti-correlation.
Note also that GCs without CN-CH anti-correlation (NGC 1851 and M22) are those with calcium abundance variations.
This also suggests that the absence of CN-CH anti-correlation is probably due to the effect of SNe enrichment, because only this mechanism can increase overall metallicity, including C, N, and heavy elements.

%%%%%%%%%%%%%%%%%%%%%%%%%%%%%%% Figure 14 'Na-O' %%%%%%%%%%%%%%%%%%%%%%%%%%%%%%%%%%%
\begin{figure}
\centering
\includegraphics[width=0.32\textwidth]{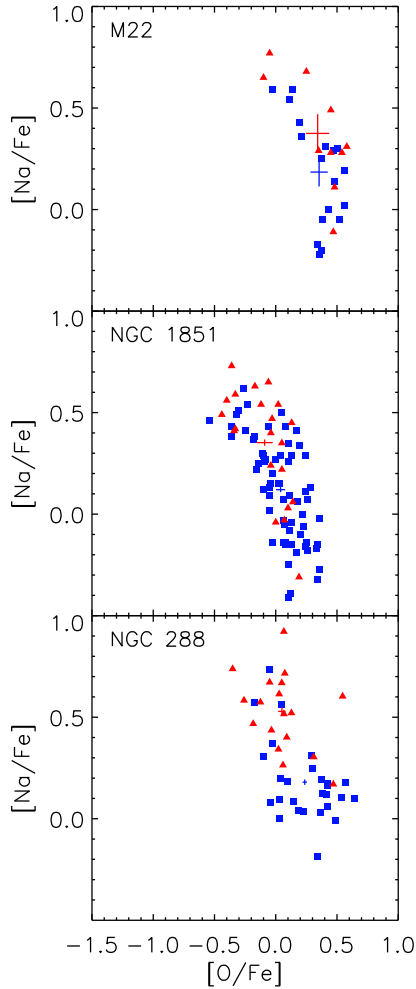}
\figcaption{The distribution of RGB stars on the Na-O plane for M22, NGC 1851, and NGC 288 \citep[data from][]{Car09b,Car11b,Mar11b}.
The redder and the bluer subpopulations are denoted by the red triangles and the blue squares, respectively.
While the difference in [Na/Fe] between the two subpopulations is significant only at 1.6$\sigma$ level for M22, NGC 1851 and NGC 288 show significant differences at more than 10$\sigma$ levels.
Mean value and the error of the mean for each subpopulation are denoted by blue and red crosses.
\label{fig_nao}}
\end{figure}
%%%%%%%%%%%%%%%%%%%%%%%%%%%%%%%%%%%%%%%%%%%%%%%%%%%%%%%%%%%%%%%%%%%%%%%%%%%%%%%%

% Section 6 (Discussion)
\section{Discussion} 
\label{sec_discussion}
We have shown that RGB stars in M22 and NGC 1851 are clearly split into two subpopulations in the narrow-band photometry employing the new Ca filter, while this is not observed in NGC 288.
From the low-resolution spectroscopy, we have confirmed that these splits in M22 and NGC 1851 are indeed originated from the difference in calcium abundance.
This confirms that a part of the RGB split reported by \citet{jwlee09a,jwlee09b} was due to the difference in calcium abundance.
In the case of CN band strength, the strong differences between the two subpopulations are found in all three GCs from our spectroscopy.
In addition, we found a systematic difference in the CN-CH correlation among these GCs.
While the well-known CN-CH anti-correlation is confirmed in NGC 288, RGB stars in NGC 1851 show no correlation between CN and CH.
For M22, the trend becomes reversed and we found a positive correlation between the two bands.
This systematic difference among three GCs is mainly due to the difference in the strength of CH band.

The origin of the presence of the two distinct subpopulations with different heavy element abundance in some GCs is not yet fully understood.
Two scenarios are suggested in the literature, all of which are apparently possible only in dwarf galaxy environment.
One is that the later generation stars in these GCs were formed in the metal enriched gas from the earlier generations (hereafter self-enrichment scenario; \citealt{Tim95}, see also \citealt{JL13}, and references therein).
The other is that these GCs formed through the merging of two proto-galactic GCs with slightly different metallicity (hereafter merger scenario; \citealt{Car10b,Car11b,BY12}).
The redder RGB stars in M22 and NGC 1851 are enhanced both in heavy and light elements.
In the self-enrichment scenario, this would suggest that the redder RGB stars were affected by SNe enrichment, together with the contamination from IMAGB stars, FRMSs, or rotating AGB stars \citep{Dec07,Dec09,VD08}.
The positive CN-CH correlation observed in M22 is also naturally explained in this scenario as the SNe enrichment would have increased both nitrogen and carbon abundances.
In the case of NGC 1851, the absence of CN-CH correlation is most likely because the enhancement in carbon abundance by SNe was not enough to overcome the depletion of the same element by IMAGB stars or FRMSs.
In this scenario, NGC 288 was not affected by SNe, but carbon was only depleted by IMAGB or FRMS.
This suggests that the systematic difference in CN and CH correlation among three GCs is caused by how strongly SNe enrichment has contributed to the chemical enrichment in GCs.
The origin of this difference probably has something to do with the difference in the initial mass of the system as more enriched gas from SNe would be retained in more massive GCs \citep[see, e.g.,][]{Bau08}.

In the case of the merger scenario, the presence of two subpopulations differing in both heavy and light elements abundances can be intuitively explained by simple merging of two GCs in the proto-dwarf galaxy environment \citep{Car10b,Car11b,BY12}.
More metal-rich GCs are, in general, also observed to be enriched in light elements \citep[see, e.g.,][]{Car09a}.
This would also explain the CN-CH correlation in M22.
The multimodal CN distribution in NGC 1851 was interpreted as an evidence of merger scenario as well \citep{Camp12}.

The Na-O anti-correlation among RGB stars, which is well established in most GCs  \citep[][and references therein]{Car09a,Car09b,Gra12a}, can provide a further constraint on these scenarios.
According to these observations, the regime occupied by stars on the Na-O plane varies with the metallicity of GC.
Therefore, stars in the GCs having similar metallicity are distributed in the similar regime on the Na-O plane.
In the merger scenario, since the difference in metallicity between the two subpopulations is small in M22 and NGC 1851, we would expect almost the same distribution on the Na-O plane.
In the self-enrichment scenario, on the other hand, the metal rich subpopulation would be placed in the Na rich regime on the Na-O plane, because the metal enriched gas used in the formation of the metal rich later generation stars would have also contaminated by AGB and/or FRMS.
Figure~\ref{fig_nao} shows the Na-O anti-correlations for the RGB stars in two subpopulations in M22 and NGC 1851, where the spectroscopic data were adopted from \citet{Mar11b} and \citet{Car11b}, respectively.
For comparison, the case of ``normal'' GCs enriched by AGB/FRMS, but not by SNe, is illustrated by NGC 288 (bottom panel of Figure~\ref{fig_nao}; data from \citealt{Car09b}), where the second generation stars are clearly different from the first generation stars in that they are enhanced in [Na/Fe] but depleted in [O/Fe] \citep[see also][]{Car09b, Pio13}.
As shown in the middle panel of Figure~\ref{fig_nao}, the metal rich RGB stars in NGC 1851 are preferentially placed in the Na rich regime \citep[see][for a similar result]{Lar12}.
The Na abundance of the metal rich group (red triangles) is enhanced by 0.27 dex in [Na/Fe] compared to the metal poor group, which is significant at 13.0$\sigma$ level.
For M22 (top panel), however, the metal poor (blue squares) and the metal rich (red triangles) RGB stars show similar distributions on the Na-O plane \citep[see also][]{Mar11b}.
The difference in [Na/Fe] between the two groups is 0.19 dex, which is significant only at 1.6$\sigma$ level.
The significant difference in [Na/Fe] between the two subpopulations in NGC 1851 would prefer the self-enrichment scenario over the merger origin.
However, for M22, the sample size is too small to reach a firm conclusion from this data set, and therefore, further observations for a larger sample of RGB stars are required.

We will continue our search for GCs originated from dwarf galaxies like $\omega$~Cen, M22 and NGC 1851.
Most of these Galaxy building block candidates are expected to be present near the galactic bulge like NGC 6388 and Terzan 5 (\citealt{Fer09,Bel13}, see also \citealt{Lee07}), where the reddening and field star contaminations are severe.
In order to overcome these obstacles, we are currently performing both narrow-band Ca photometry and low-resolution multi-object spectroscopy for these GCs.
As will be demonstrated in our forthcoming paper, combining these two modes of observation makes it possible to effectively disentangle the effects from differential reddening and metallicity.

% Acknowledgement
\acknowledgments We are grateful to the anonymous referee for a number of helpful suggestions. We also thank the staff of LCO for their support during the observations. Support for this work was provided by the National Research Foundation of Korea to the Center for Galaxy Evolution Research (No. 2010-0027910) and by the Korea Astronomy and Space Science Institute under the R\&D program (Project No. 2014-1-600-05) supervised by the Ministry of Science, ICT and future Planning. C. I. J. gratefully acknowledges support from the Clay Fellowship, administered by the Smithsonian Astrophysical Observatory.\\

% References

\end{document}